\documentclass[preprint,authoryear,12pt, review]{elsarticle}

 \usepackage{graphicx}

\usepackage{amssymb}

 \usepackage{lineno}

\journal{Icarus}

\begin{document}

\begin{frontmatter}
%
%
%
\title{Field-aligned beams and reconnection in the jovian magnetotail}
%
%
\author[a1]{E. A. Kronberg\corref{cor}}
\ead{kronberg@mps.mpg.de}
\author[a2]{S. Kasahara}
\ead{kshr@stp.isas.jaxa.jp}
\author[a1]{N. Krupp}
\ead{krupp@mps.mpg.de}
\author[a1]{J. Woch}
\ead{woch@mps.mpg.de}
\address[a1]{
Max-Planck-Institut f\"{u}r Sonnensystemforschung, Max-Planck Str., 2, Katlenburg-Lindau, 37191,
Germany.}
\address[a2]{
Institute of Space and Astronautical Science, JAXA, 3-1-1 Yoshinodai, Chuo-ku,
Sagamihara, Kanagawa 252-5210, Japan.}
\cortext[cor]{Corresponding author, Tel. +49 5556 979481, Fax. +49 5556 979240}
\begin{abstract}
The release of plasma in the jovian magnetotail is observed in the form of
 plasmoids, travelling compression regions, field-aligned particle beams and
flux-rope like events. We demonstrate that electrons propagate along the
magnetic field lines in the plasma sheet boundary layer (PSBL), while close
to the current sheet center the electron distribution is isotropic. The
evidences of the counterstreaming electron beams in the PSBLs are also
presented. Most of the field-aligned energetic ion beams are associated with
the field-aligned electron beams and about half of them have the bipolar
fluctuation of the meridional magnetic field component. Moreover they often
show a normal velocity dispersion for the different species which fits well
in the scenario of particle propagation from a single source. All features
above are observed during jovian reconfiguration events which are typically
bonded with plasma flow reversals. From all these characteristics, which are
based on energetic particle measurements, we believe that the reconfiguration
processes in the jovian magnetotail are associated with reconnection.
\end{abstract}
\end{frontmatter}
\section{Introduction}
One of the dynamic reconfiguration processes of the magnetosphere is called
substorm and is a key process responsible for the energy transport and
release in the terrestrial magnetosphere. It is believed that reconnection is
responsible for the break of stability in the magnetospheric configuration
and also for the transfer from magnetic to kinetic energy, e.g.
\citep{Angel08}. During this process the magnetic field forms a so-called
X-line surrounded by the diffusion region, where acceleration of particles
takes place. In the diffusion region the electric field accelerates
high-energy ions and electrons away from the magnetic reconnection region and
electrons at low-energy flow toward the reconnection site simultaneously
\citep{Nagai01, Manapat06}. The energetic particles are released as bursty
bulk flows and are widely studied by e.g. \citet{Angel94}. The bursty bulk
flows often contain structures with closed magnetic field lines -- plasmoids.
Such a large tailward-moving loop-like magnetic structure is generated as the
result of X-type reconnection on the closed field lines in the near
magnetotail \citep{Hones79}. In two dimensions the typical observational
signatures of a plasmoid have been defined by a bipolar magnetic deflection
in the meridional direction. The examination of the three-dimensional
topology and morphology of plasmoids in the Earth's magnetotail showed rather
complicated configurations compared to the original 2-dimensional Hones
scenario: helical flux ropes and travelling compression regions (TCRs)
detected as moving magnetic imprints of a plasmoid in the lobes
\citep{Zong04,Slav95,Slav03a}. A tailward moving plasmoid in the central
plasma sheet temporarily compresses the magnetic flux in the tail lobes
because the cross-sectional area of the lobe is reduced. In the area between
the central plasma sheet and lobe a layer of particles moving parallel to the
magnetic field arises from a rapid change of the flux tube volume outside the
diffusion region \citep{Schindler07}. However, these field-aligned beams
could also be of a different origin than reconnection, e.g. due to Fermi
acceleration \citep{Vog06, Grig09}. Thus in the terrestrial magnetosphere two
distinct
groups of field-aligned beams can be derived:\\
(1) field-aligned ion beams
accompanied with field-aligned electrons and associated with the X-line
reconnection;\\
(2) ion field-aligned beams accompanied by isotropic electrons
accelerated by the quasi-steady dawn-dusk electric field.
Substorm-like (or) reconfiguration processes have been observed at Jupiter by
the particle and fields instruments onboard the Galileo spacecraft
\citep{Krupp98, Woch98,Woch99, Kron05}. It is believed that the
reconfiguration processes in the jovian magnetosphere are rather internally
driven by ion mass loading from the moon Io and fast planetary rotation,
although we cannot fully exclude the influence of the solar wind
\citep{Kron09b}. This internal driving mechanism leads to a periodic release
of plasma in the magnetotail with a repetition period of about 3 days. The
thinning of the plasma sheet prior to the mass release reminds the growth
phase of the terrestrial substorm \citep{Ge2007}. This plasma sheet
reconfiguration due to internal energy loading leads to a formation of the
X-line in the jovian magnetotail at approximately 80 $R_J$ \citep{woch02,
Vogt10}. The scale size of downtail released plasmoids was estimated to be
about 9 $R_J$ \citep{Kron08b}.
Reconnection processes at Jupiter seem to play an important role in the
reconfiguration process as reported by
\citet{Nishi83,Russ98,Russ00,Vogt10,Ge2010}. Reconnection conditions in the
jovian magnetotail are satisfied just before the energetic particle release
\citep{Zim93,Kron07}. Statistical studies of bursty bulk flows and plasmoids
were carried out by \citet{Kron08b, Vogt10}. Also data from the New Horizons
spacecraft during the Jupiter flyby showed evidence of periodic plasma
injections possibly caused by reconnection \citep{McNutt07,McComas07}. Some
field-aligned beams were discussed in the substorm-like context by
\citet{Woch99}.
In this paper, we show for the first time observations of a TCR and plasmoids
together with field-aligned beams and consider their features more detailed
in the context of X-line reconnection using data from the Galileo Energetic
Particle Detector (EPD) \citep{Will92} and from the magnetometer (MAG)
\citep{Kiv92}. The paper is organized as follows: In Section \ref{Sec:1} we
introduce shortly the instruments onboard Galileo used to identify
field-aligned beam features in the measurements. In Section \ref{scene} we
present a typical jovian reconfiguration event with field-aligned beams and
dynamic processes preceding and following the beam. In Sections
\ref{timing}-\ref{event:4} we show examples of four field-aligned beams and
their wide spectrum of characteristics. Section \ref{discussion} discusses
results and provides a statistical analysis. Section \ref{summary} summarizes
the observations.
\section{Instrumentation}\label{Sec:1}
The EPD was an instrument onboard the Galileo spacecraft designed to measure
the characteristics of the charged particle population such as energies,
intensities, ion composition and angular distribution to determine, in
particular, the configuration of the jovian magnetosphere \citep{Will92}. The
advantages of the EPD instrument compared to similar instruments on previous
missions were the $4\pi$ steradian angular coverage for jovian energetic
particles and the extended coverage of particle energies.
EPD consisted of two double-headed detector telescopes: the Low Energy
Magnetospheric Measurement System (LEMMS) and the Composition Measurement
System (CMS). CMS measured the ion fluxes with discriminating ion species
using the time-of-flight technique. Its energy ranges were 80 to 1250 keV
(protons), 27 keV/nuc to 1 MeV/nuc (helium ions), 12 to 562 keV/nuc (oxygen
ions), and 16 to 310 keV/nuc (sulfur ions). On the other hand, LEMMS observed
the electron flux in the energy range between 15 and 11000 keV and the
species-integrated ion flux between 22 and 12400 keV. In this case electrons
and ions were separated by a permanent magnet. Although these telescopes have
narrow field-of-view (the full opening angles are 18$^\circ$ and 15$^\circ$
for CMS and LEMMS, respectively), an almost full-sphere coverage is achieved
by the motion of a turntable combined with the spacecraft spin (the exception
is a small solid angle ($\le$ 0.1 sr) along the spin axis which is blocked to
avoid direct sun light in the detector). The unit sphere is divided into 16
(see Figure 6.2, in \citep{Lagg98}) or 6 sectors dependent on the energy
channel. To study the flow direction we analyze 16-sector resolution data in
this work. Data are accumulated within sampling times of $\sim$11.5 min (for
this study) and allocated to the angular sectors. Each sector is $\sim
\pm$45$^\circ$ wide. In the record time mode EPD/LEMMS count rates are
sampled every 1.25 seconds over up to 64 sectors. The description of LEMMS
and CMS energy channels used in our analysis one can find in Table
\ref{t2le}.
The first order anisotropies were calculated using technique described by
\citet{Krupp01}, and references therein.
The pitch angle distributions (PAD) are presented in the detector frame --
Detector Pitch Angles (DPA). The flow is parallel to the magnetic field when
DPA is close to 180$^{\circ}$ and respectively anti-parallel when DPA is
close to 0$^{\circ}$.
Particle pitch angles are determined with the magnetic field data obtained by
the Galileo magnetometer. We discuss pitch angles only when the magnetic
field direction varies insignificantly during a sampling time, since
otherwise the pitch angle determination is unreliable.
The magnetic field observations have a time resolution of $\sim$24 s, which is
 much higher than that of the particle measurements. This does not affect our
 results, however, as we compare the
observed field-aligned beams in the particle data with features in the
magnetic field and not the other way around. Still, it is very likely that we
miss many events with time scales less than 11 min in the EPD data set.
\section{Observations}
\subsection{An overview event}\label{scene}
\begin{figure}[h]
 \includegraphics[width=12.0cm]{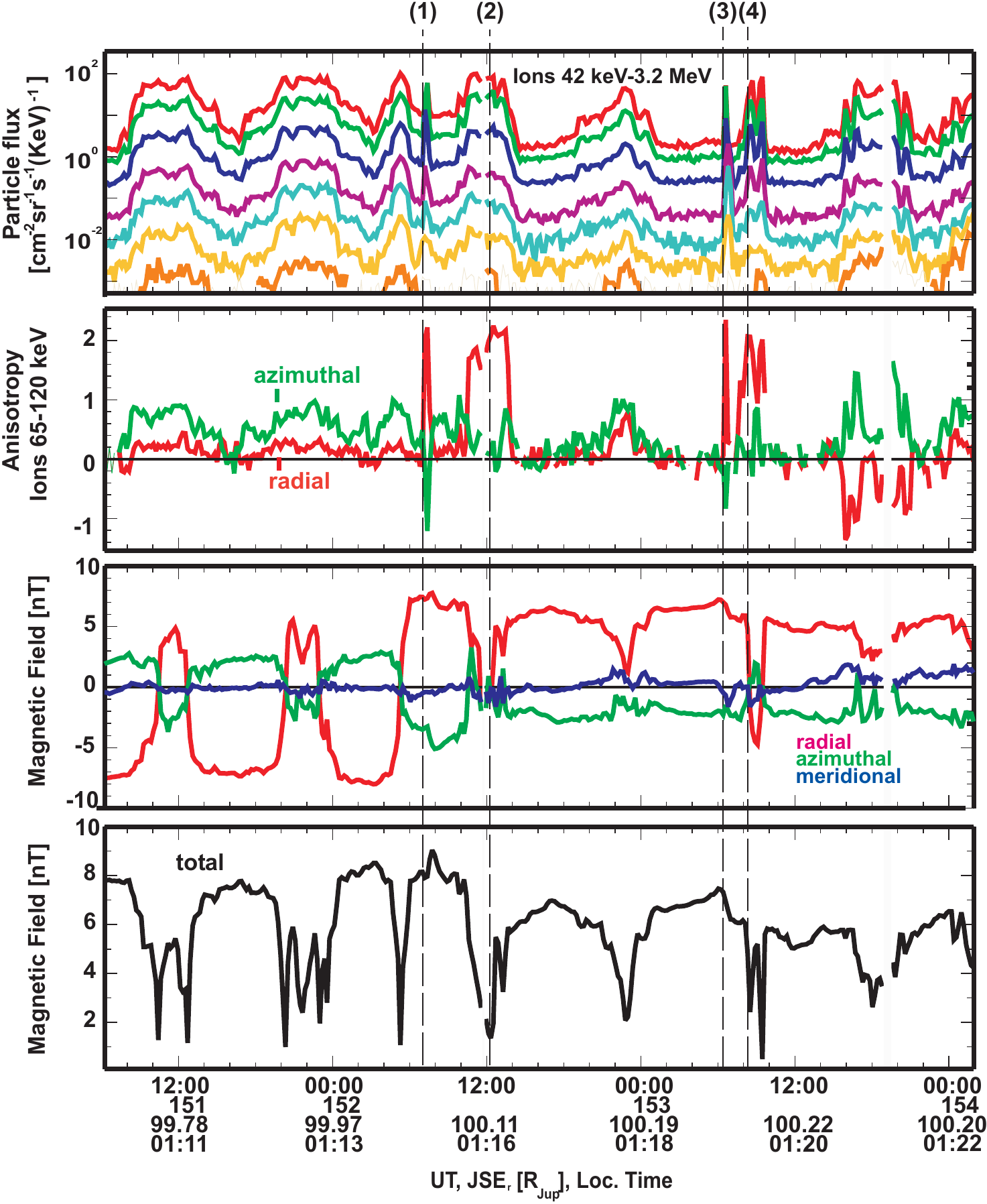}
 \caption[Energetic particle and magnetic field observations on Galileo orbit G8
 from DOY 151, 0600 to DOY 154, 0200 in 1997.]
 {Energetic particle and magnetic field observations on Galileo orbit G8
 from DOY 151, 0600 to DOY 154, 0200 in 1997. From top to bottom are
  displayed: (first panel) omnidirectional ion intensities for different
 energy channels at 0.042-3.2 MeV (from a1 to a7, see Table \ref{t2le});
 (second panel) first order ion directional flow anisotropies in the radial
 (positive is outward) and corotational direction; (third panel) the magnetic
 field components in SIII coordinates and (fourth panel) its magnitude. The
 vertical dashed lines denote the field-aligned beams and the bursty bulk
 flows.} \label{15}
 \end{figure}

A typical substorm-like or reconfiguration event in the jovian magnetosphere
is shown in Figure \ref{15}. The figure shows the time interval from day 1997
151, 06:00 to day 154, 02:00. During this time Galileo was located in the
midnight sector of the jovian magnetotail (0116 Local Time (LT)) at about a
radial distance R = 100 $R_J.$ The upper panel displays omnidirectional ion
intensities in seven energy channels covering the energy range between 22 keV
(a1 channel) and 3.2 MeV (a7 channel). The second panel from the top shows
the radial (red) and azimuthal (green) component of the ion directional flow
anisotropy as derived from measurements of 65-120 keV ions. The lower two
panels present the magnitude and components of the measured magnetic field in
the SIII-system. In this system the radial magnetic field component is
positive in the radially outward direction, the azimuthal component is
positive in the direction of Jupiter's rotation and the south-north
(meridional) component is positive southward \citep{Des83}.
First signatures of a mass-release process are observed at day 152, 07:00
with an unusual (for the lobe region) increase of the ion intensities (see
vertical dashed line (1) in Figure \ref{15}). The corotational flow, typical
for the jovian magnetosphere (see the ion anisotropy denoted by the green
line in the second panel from the top) is disrupted by a strong tailward
(radially outward) directed anisotropy spike or an field-aligned ion beam (the
red line) in the lobe / PSBL (that the beam is field-aligned will be proven
later). This is followed by the bursty flow ion anisotropy event or bursty
bulk flow (denoted by the vertical dashed line (2)) which arrives with
bipolar features in the south-north magnetic field component at day 152,
10:00 when Galileo approached the vicinity of the
 current sheet center. About 16 hours later another group of tailward
 directed bursts with similar characteristics arrives (see vertical dashed
 lines (3) and (4) in Figure \ref{15}). Then in the subsequent plasma sheet
 encounter radially inward directed bursts (the negative red spikes) are
 observed implying that the X-line moved tailward over the spacecraft. The
observations of the flow reversals are very typical for the jovian
magnetotail, see e.g. \citet{Kron05, Kron09b}. It is interesting that just
before the bursty flows the plasma sheet changed its displacement relative to
the spacecraft from being more in the southern part of the plasma sheet to
being in the northern part of the plasma sheet almost all of the presented
time (according to the radial magnetic field component behavior). Such plasma
sheet displacements were discussed by \citep{Vas97,Waldrop05}. They suggest
that these displacements could be associated to a change in the solar wind
conditions.
 More details on the reconfiguration or substorm-like events can be found in
 \citet{Woch99, Kron05, Kron08b}. This example is shown to present the global
 picture of the magnetotail dynamics before going into details.
\subsection{Event 1: Counterstreaming field-aligned electrons and possible interpretation}\label{timing}
\begin{figure}
\includegraphics[width=12cm]{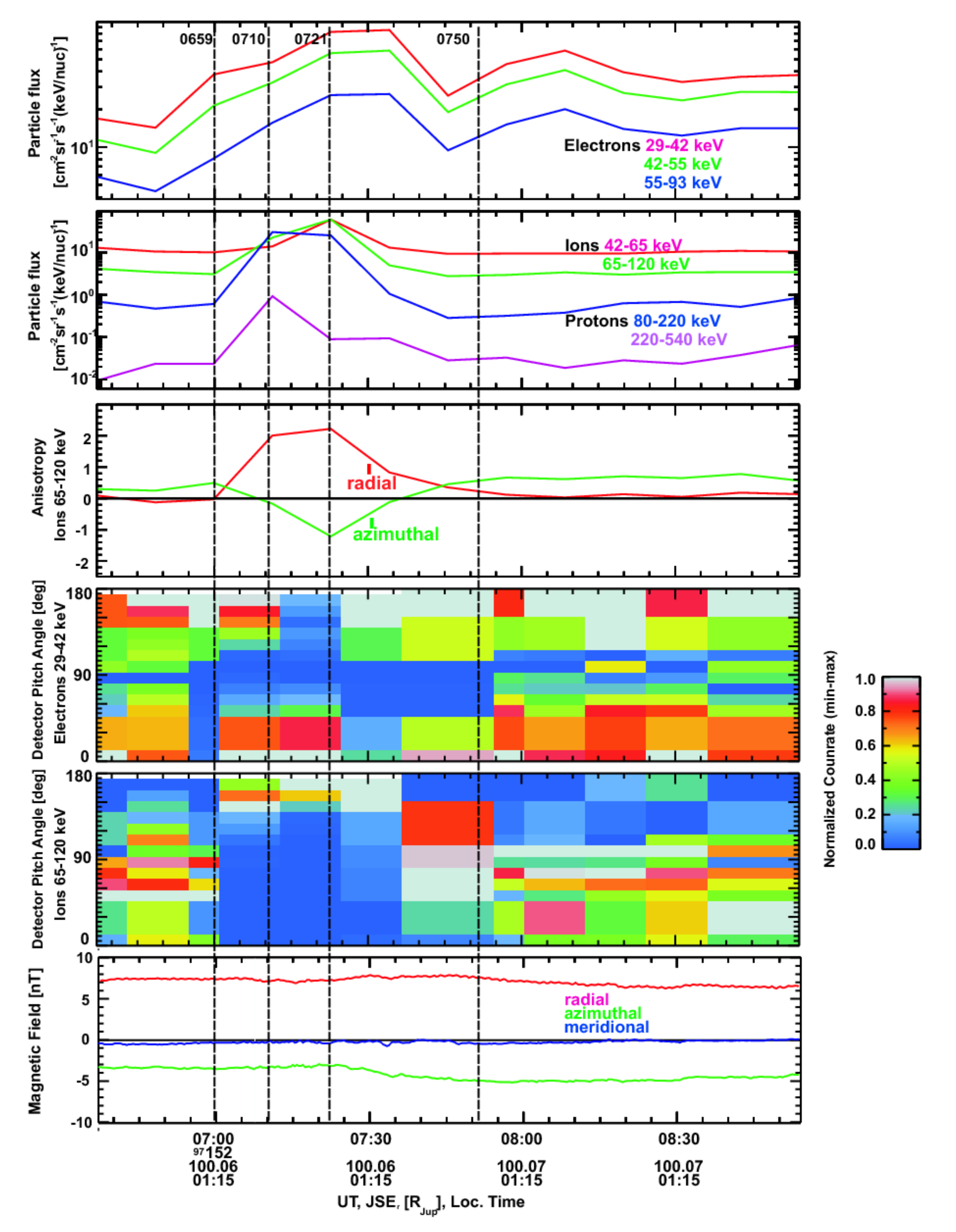}
\caption{Event 1, DOY 152 of 1997, 0637-0854 UT: The plot includes from top to bottom:
 electron intensities at 29-304 keV (first panel); ion and proton intensities
at 42 keV-1.7 MeV and 80-540 keV, respectively (second panel); the
first-order ion anisotropies at 65-120 keV (third panel); DPA electron
distributions at 29-42 keV (fourth panel) and for the ions at 65-120 keV
(fifth panel). Sixth panel shows magnetic field data plotted in the SYS-III
coordinate system: red is $\sim B_r$, green is $\sim B_{\phi}$ and blue is
$\sim B_{\theta}.$
 }
\label{152}
\end{figure}

\begin{figure}
\includegraphics[width=16.6cm]{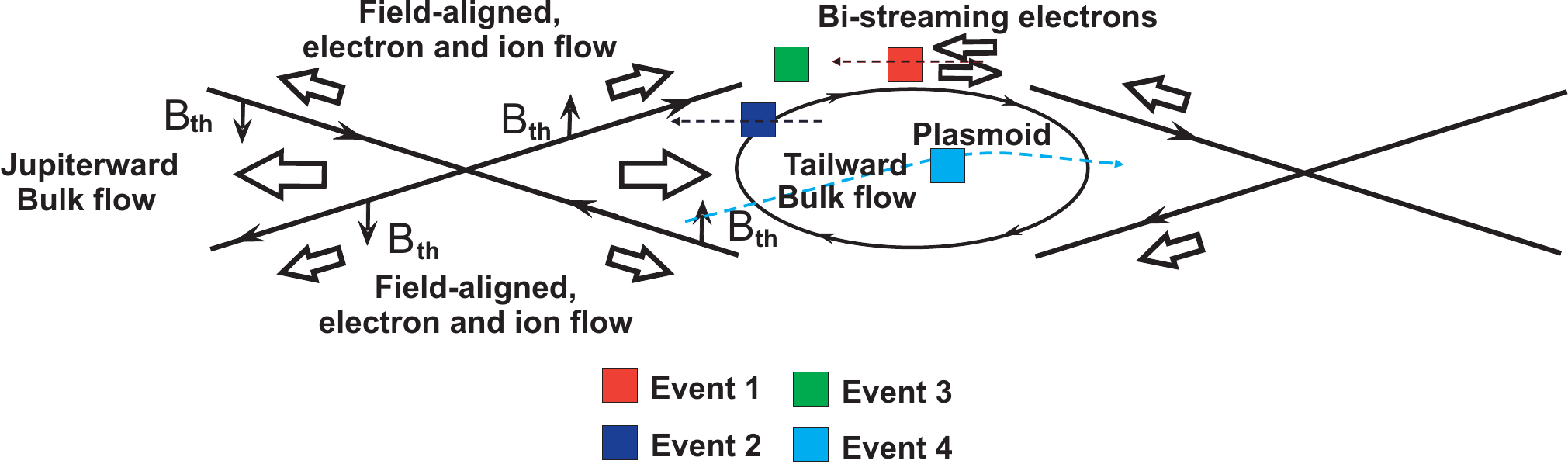}
\caption{Sketch of the events locations. Dashed arrows show roughly Galileo trajectory.
 }
\label{scetch}
\end{figure}

One example of a field-aligned beam during reconfiguration events is shown in
Figure \ref{15}. It was recorded at 100 $R_J$ away from the planet in the
magnetotail, on day 97 152, 06:30-09:00. The detailed particle and the
magnetic field observations for this event are presented in Figure \ref{152}
and the location relative to the plasma sheet configuration is sketched
in Figure \ref{scetch}.
According to the strength and direction of the magnetic field the spacecraft
was located in the north lobe/or PSBL. First the electrons arrive with
energies from 15 to 304 keV (see electron intensities in Figure \ref{152},
panel 1) at 0659 UT. These electrons were flowing parallel to the magnetic
field in the lobe region (see the increase in electron intensities close to
180$^{\circ}$ in DPA distributions, panel 4). After 0659 UT electrons moved
bi-directionally parallel to the magnetic field (see the increase in
intensities close to 0$^{\circ}$ and 180$^{\circ}$). Between 0710 UT and 0750
UT the observed electrons changed their direction of motion from being
anti-parallel 0710-0721 UT to mainly parallel 0721-0750 UT relatively to the
magnetic field. At 0750 UT the electron distribution becomes isotropic.
This field-aligned electron beam is associated with the reconfiguration event
from the list published by \citet{Kron05}. After the field-aligned beam was
observed the plasma sheet center moved towards Galileo (see Section
\ref{scene}, Figure \ref{15}). As it is expected from the reconnection
scenario the bulk release of particles (see vertical dashed line (2) in
Figure \ref{15}) at the Alfv\'{e}nic speed is observed at 1200 UT on day 152,
as studied by \citet{Kron08b}. From the comparison of the flow speed of the
bursty bulk flows was compared with the Alfv\'{e}n speed. We can assume that
during this time the reconnection process was ongoing and that the electron
beam has possibly been mirrored at the next X-point as in a magnetic bottle
bouncing between two mirror points on loop field lines (i.e. multiple X-line
formation is assumed, see also Kasahara et al., submitted to JGR) or launched
from another X-line. Also the bi-directional streaming electrons are commonly
observed in the terrestrial plasma sheet boundary layers \citep{Walsh11}.
Ions arrived at least 11 min later at 0710 UT. The third panel of Figure
\ref{152} shows the first order directional anisotropy components of ions
(65-120 keV) in radial and azimuthal directions. During this event the ions
moved clearly predominantly radially outward with an anti-corotational
component. Those ions moved also along the magnetic field direction as
indicated in the pitch angle distributions (panel 5 of Figure \ref{152}). Ion
fluxes exhibit an energy dispersion in the sense that the intensities of high
energy ions increase first.
Assuming that the electrons have a single source (e.g. released at the
reconnection site during the onset phase), the distance from Galileo
to the source (e.g. the reconnection line) can be roughly estimated from the
delay time between different species. We must point out that the time
resolution of the EPD data is 11 minutes and therefore the following
calculations have large error bars. Ions measured in the a1 channel (42-65
keV), mainly protons, with a geometric mean energy of 52 keV (speed of $\sim$
3200 km s$^{-1}$) arrive $\sim$ 22 min later than electrons. This gives a
distance of 60 $ R_J \pm $ 30 $R_J.$ According to these calculations the
distance from the source to Galileo was between 30 $R_J$ and 90 $R_J$.
Consequently a source would be located at $\sim$ 40 $ R_J \pm $ 30 $R_J$ from
the planet. The source location is slightly closer to the planet than those
derived by \citet{woch02} but still similar within the error bar. Even if the
delay calculations are rough, this event presents an energy dispersion of
different particles: electrons arrive first, and then high energy ions
followed later by lower energy ions. This implies a single source of these
field-aligned beams.
One can claim that the observed dispersion is related to gyroradius effect or
gradient effects. We estimate the gyroperiod of the proton of $\sim$9 s in
this magnetic environment. We see that the accumulation time of 11 min is
much longer than the proton gyroperiod. Additionally, the spacecraft passes
0.002 $R_J$ during one accumulation time of the EPD instrument (the Galileo
speed is $\sim$3.2$\cdot$10$^{-6}$ $R_J/s$). This distance is the same order
of magnitude as the proton gyroradius of a field-aligned beam. Therefore, the
spacecraft could be considered basically stationary relative to the ion
motion. The ion dispersion is also not related to $\vec{E}\times \vec{B}$
drift, as they would lead to much longer time delays.
\subsection{Event 2: TCR and magnetic-islands like structure}
\begin{figure}
\includegraphics[width=11cm]{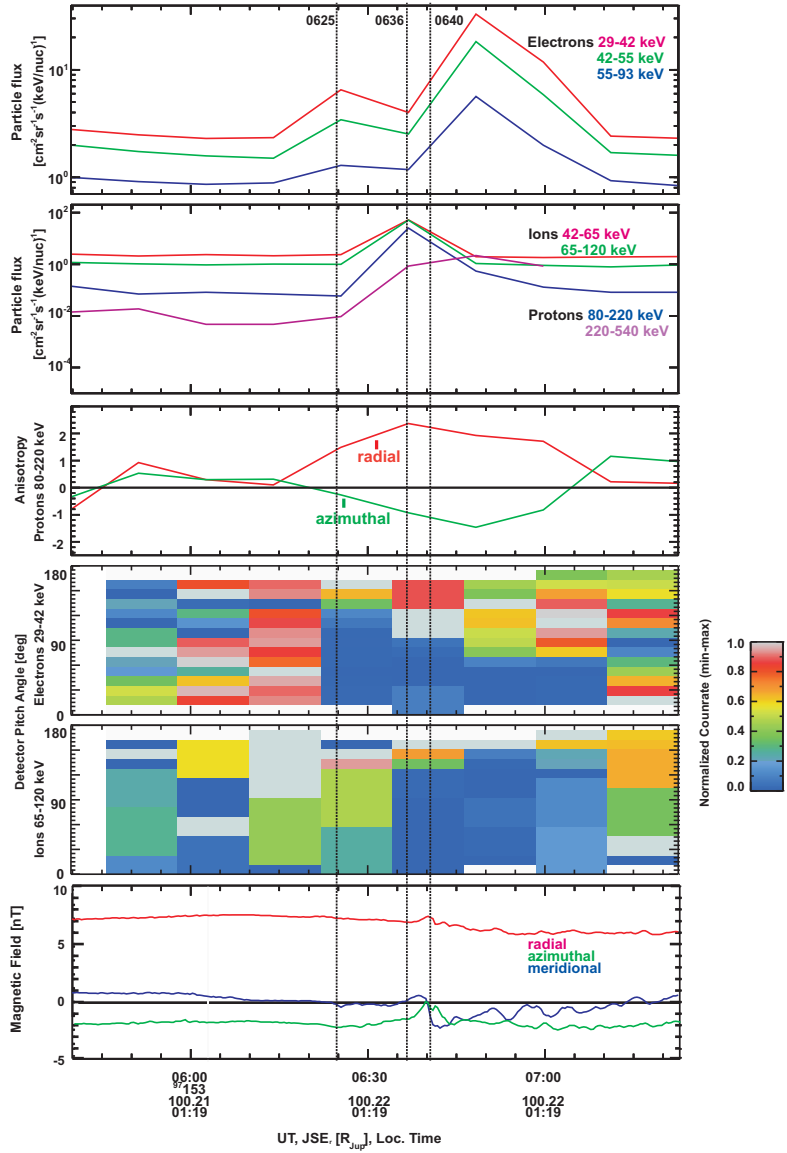}
\caption{Event 2, DOY 153 of 1997, 0539-0722 UT: The plot includes from top to bottom: the
 electron intensities (first panel), ion and proton intensities (second
 panel), the first order ion anisotropies at 65-120 keV (third panel), DPA
electron distributions at 29-42 keV (fourth panel) and for the ions at 65-120
keV (fifth panel). Sixth panel shows magnetic field data plotted in the
SYS-III coordinate system: red is $\sim B_r$, green is $\sim B_{\phi}$ and
blue is $\sim B_{\theta}.$
 }
\label{153}
\end{figure}

Another field-aligned beam was observed during the reconfiguration event on
DOY 153 at 0620 and is shown by vertical dashed line (3) in Figure \ref{15}.
Details are presented in Figure \ref{153} in the same format as for Event 1
in Figure \ref{152}. During this beam significant increases by more than one
order of magnitude of the ion intensities (42 keV to 1.7 MeV) and by one
order of magnitude of the electron intensities (15 to 304 keV) are observed.
Again, the first order flow anisotropy shows that the ions during this event
were flowing radially outward in anti-corotational direction. Location of
this event relative to the plasma sheet configuration is sketched in Figure
\ref{scetch}.
The meridional magnetic field component $B_{\theta}$ becomes negative at 0625
UT (northward directed) suggesting a tailward flow in the plasma sheet
\citep{Kron08b} in case of a neutral line formation. At the same time the
field-aligned electrons followed by the field-aligned ions (arrived later at
0636 UT) are observed. The delays between electrons and ions are the same as
in the example in Section \ref{timing} suggesting similar nature of the
field-aligned beam. At $\sim$0640 UT a signature of a flux rope or TCR is
seen in the magnetic field, i.e. increase in the radial magnetic field
component, bipolar signature in the meridional magnetic field component.
Additionally, the disappearance of the anti-phase signature between the
azimuthal and radial magnetic field components as in usual quiet
configuration ($\alpha =\tan^{-1}(B_{\phi}/B_r)$ reaching zero) is observed.
The flux rope-like magnetic field (increased radial magnetic field component
together with the bipolar signature in the meridional magnetic field
component, see e.g., \citet{Zong04}) is associated with ion field-aligned
beam parallel to the magnetic field. The proton intensity increased by almost
3 orders of magnitude. After the flux-rope structure decaying oscillations in
the meridional magnetic field component are observed. These oscillations
could indicate small scale magnetic islands (probably created by magnetic
reconnection), or they are imprints of the post-plasmoid plasma sheet
structure as shown in \citet{Mukai00}. We conclude that these field-aligned
electron and ion beams are (1) associated with TCR mentioned as a
reconnection event in the study of \citet{Vogt10}, (2) followed by the bursty
bulk flow reversal (see vertical line (4) in Figure \ref{15}), and (3) show
the energy dispersion. All these features imply that the observed
field-aligned electron and ion beams are associated with a reconnection
process.
\subsection{Event 3: Observations of bi-directional field-aligned electron flow
-- timing and reconnection scenario}\label{sec:283}
\begin{figure}
\includegraphics[width=10cm]{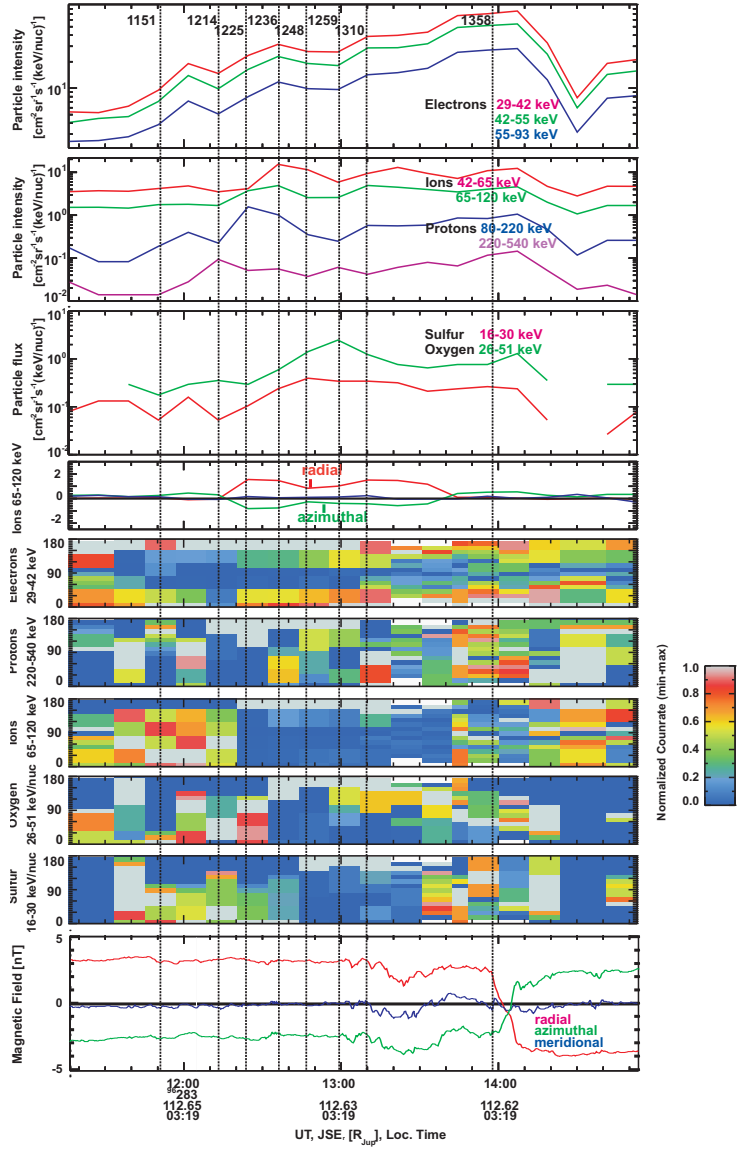}
\caption{Event 3, DOY 283 of 1996, 1116-1453 UT: The plot includes from top to
bottom: the electron intensities (first panel), ion and proton intensities
 (second panel), sulfur and oxygen intensities (third panel), the first order
 ion anisotropies at 65-120 keV (fourth panel), DPA distributions of
electrons at 29-42 keV (fifth panel), protons at 220-540 keV (sixth panel),
ions at 65-120 keV (seventh panel), oxygen at 26-51 keV/nuc (eighth panel)
and sulfur at 16-30 keV/nuc (ninth) and the magnetic field components as in
previous Figure (panel ten).
 }
\label{283}
\end{figure}

Another example of field-aligned particle beams is shown in Figure \ref{283}
observed on day 1996 283 from 1100 to 1500 UT . During this time Galileo was
located in the pre-dawn sector of the magnetotail (03:19 LT) at R = 113
$R_J.$ Location of this event relative to the plasma sheet configuration is
sketched in Figure \ref{scetch}. These field-aligned beams preceded the
bursty bulk flows in the reconfiguration event from the list by
\citet{Kron05}. The first order flow anisotropy shows that the ions during
this event flow radially outward in anti-corotational direction. Here again
the dispersion is observed. At $\sim$1151 UT, a bi-directional, field-aligned
electron flow (29 to 42 keV) arrival was observed (seen in the DPA
distribution). Then at $\sim$1214 UT the electron flux was mono-directional
field-aligned and changed to bi-directional field-aligned again between
$\sim$1225 and $\sim$1259 UT. Simultaneously with this strong electron beam
(in our time resolution) protons with energies from 220 to 540 keV arrived at
$\sim$1214 UT. At $\sim$1225 UT the field-aligned ions with energies of
65-120 keV and protons (80 to 220 keV) followed by oxygen (26 to 51 keV/nuc)
arrived at $\sim$1248 UT and eventually sulfur (16 to 30 keV/nuc) at
$\sim$1236 UT.
Assuming that the X-line as the source of all these flows, this scenario can
be checked. As the arrival of the electrons is not clearly determined, we
estimate the distance X to the X-line using the formula $\Delta t=X/v_s-
X/v_p,$ where $v_s$ is the sulfur speed and $v_p$ is the proton speed,
$\Delta t$ is the time delay between the arrival of protons and sulfur
$\sim$23 min. We obtain the distance from Galileo to the X-line equal to
52$\pm$ 26 $R_J$. This also agrees with the statistical X-line derived by
\citet{woch02}. For the given location it is approximately at 70 $R_J$.
Therefore, it is likely that particles were released from a single extended
source and the time-of-flight dispersion theory works and the presented
example fits the X-line reconnection scenario. It is not likely that
gyroradius effects and gradient effects influence the results for this event,
as the magnetic field during the field-aligned beam was stable and the
spacecraft was approximately at the same position in the lobe region. We also
note that this event was not considered as a reconnection event by
\citet{Vogt10}.
\subsection{Event 4: The structure of an electron beam}\label{event:4}
\begin{figure}
\includegraphics[width=12cm]{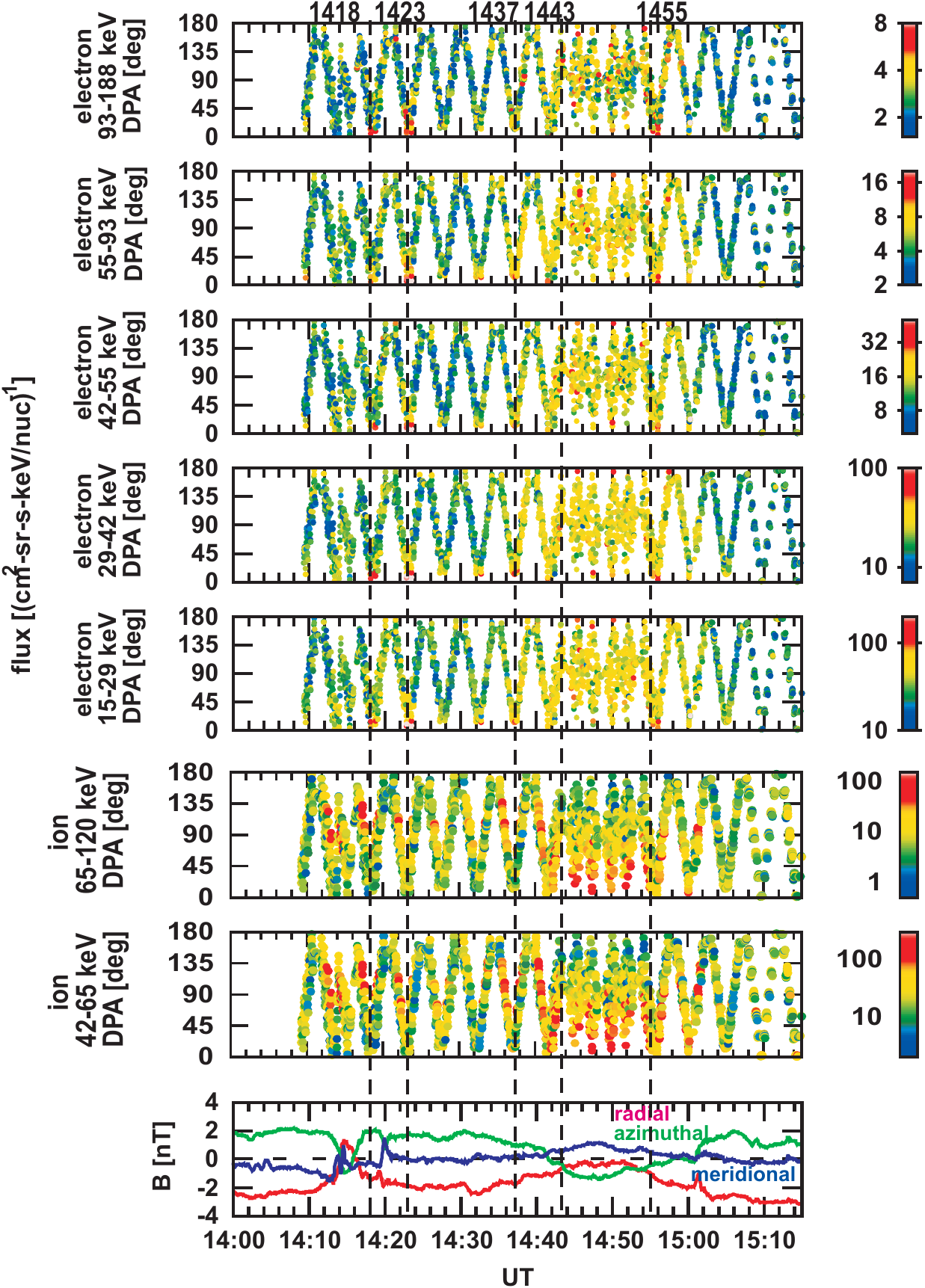}
\caption{Event 4, DOY 235 of 1997, 1400-1515 UT: The plot includes full
resolution DPA distributions for electrons at energies 15-188 keV (first
 to fifth panels), ions at energies 65-120 keV (sixth panel) and 42-65 keV
(seventh panel). The magnetic field components are as in previous Figures
(sixth panel). } \label{hr}
\end{figure}

Figure \ref{hr} presents an event on DOY 235 of 1997 showing the spacial
structure of the electron beam. During this event Galileo was located in the
pre-dawn sector of the magnetotail (00:56 LT) at R = 129.9 $R_J.$ As the
particle data were acquired with the full time resolution in this particular
period, we see the covered pitch angle and the associated particle flux due
to the instrument rotation. The magnetic field is plotted in SIII coordinate
system. Galileo was located southward of the current sheet center during most
of this period.
Anti-parallel field-aligned electron beam is observed in several channels (15
to 188 keV), from $\sim1418$ to $\sim1437$ UT (see in DPA distributions as
red dots close to 0$^{\circ}$). The beam almost disappeared during 1427-1429
UT, indicating that this is the intermittent and/or localized beam. This beam
is associated with the bipolar signatures in the meridional magnetic field
component at $\sim$1420 UT. This event shows a mono-directional electron
beam, consistent with an X-line planetward of the Galileo spacecraft.
The field-aligned ion beam is seen at energies 65-120 keV and 42-65 keV
during 1443-1455 UT. During the neutral sheet approaching at around 1443-1455
UT, electron flux is enhanced by a factor of $\sim$2 compare to the period at
$\sim$1420-1435 UT in the plasma sheet boundary layer, the particle
distribution is isotropic. The fluxes in the plasma sheet are higher compared
to those at the boundary layer. All these suggests a closed field line
topology \citep{Sarafop97}. An unambiguous evidence of the beam layer
structure is seen. Namely, the electron beam propagates at the outer edge of
the thin current sheet and an isotropic distribution is observed around the
current sheet center. This cannot be clearly evidenced by low-resolution data
due to the ambiguous pitch angle determination around the magnetic field
reversal.
The field-aligned beam is preceded by the plasmoids and associated with the
magnetic field dipolarization. The meridional magnetic field component is
already negative before (unfortunately the particle data are not available in
that resolution before $\sim$1408 UT). Therefore, there could be a formation
of a large plasmoid or post plasmoid plasma sheet followed by smaller
asymmetric plasmoid (at $\sim$1414 UT) and dipolarization (at $\sim$1420 UT)
characterized by a strong and steep increase of the meridional magnetic field
component, preceded by a much less negative dip, as in the terrestrial cases
observed and simulated by e.g. \citet{Sitnov09, Runov09,Ge2011}. This bipolar
change of the meridian magnetic field component followed by a dipolarization
front can also be explained to be caused by the tailward movement of
reconnection site, followed further by dipolarization of the magnetic field
at planetward side of reconnection site. During the plasmoid passage, the
neutral sheet crossing took place during between 1414 and 1416 UT and no
electron flux enhancement was observed. This is different from the
terrestrial case where the plasmoids centers are associated with increases in
the electron flux \citep{Chen08}. An approximate location of the spacecraft
trajectory relative to the plasma sheet configuration, between 14:00 and
14:18 UT, is sketched in Figure \ref{scetch}.
\section{Discussion}\label{discussion}
\begin{figure}
\includegraphics[width=10cm]{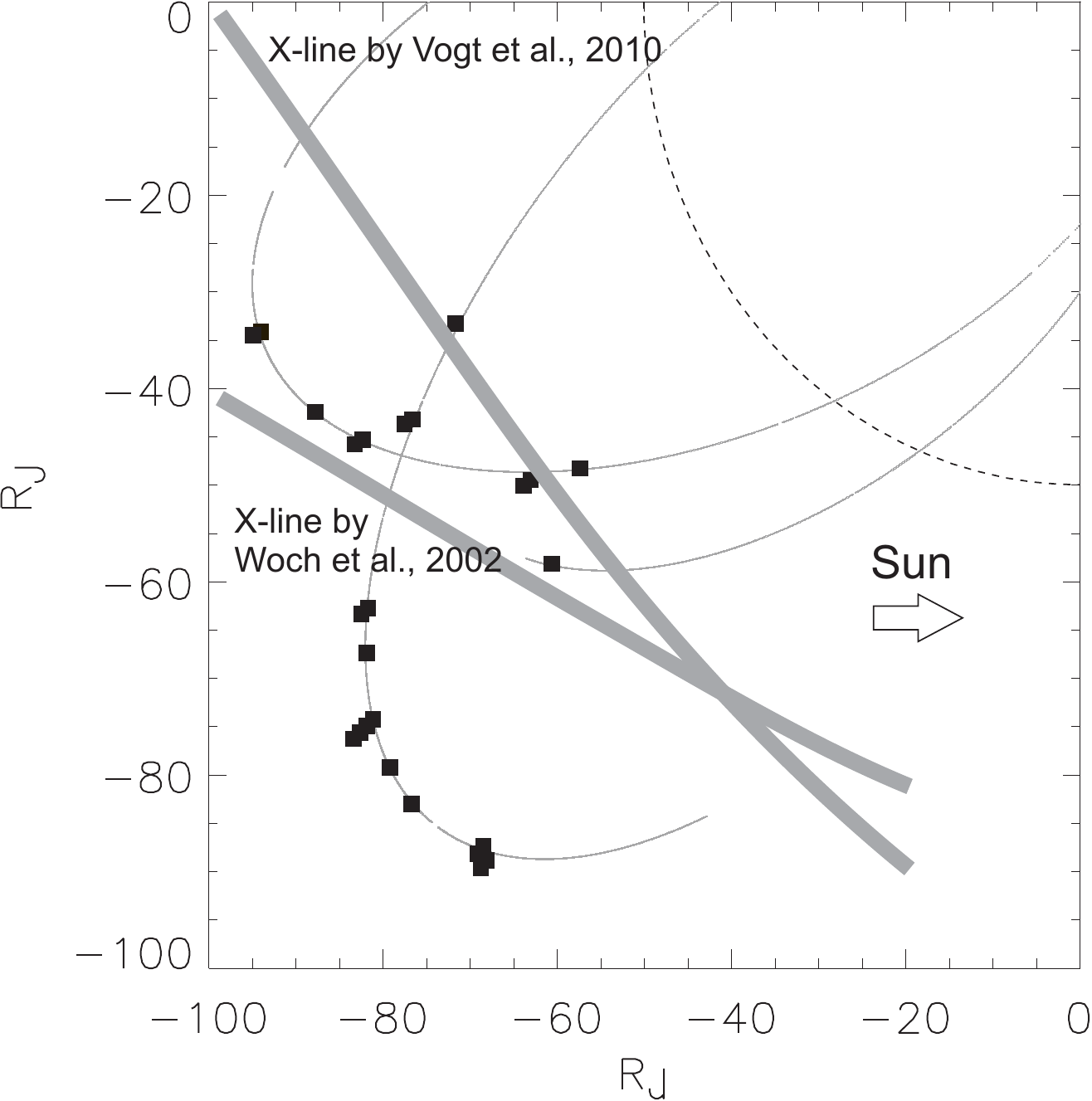}
\caption{The location of the field-aligned beams (black squares) in the
Jupiter Solar Ecliptic coordinates, in projection into the equatorial plane.
The thin grey lines show the Gaileo trajectories of G2, E6 and G8 orbits.
Two thick grey lines denote the approximate location of the X-lines derived
by \citet{woch02} and \citet{Vogt10}.} \label{map}
\end{figure}

\begin{figure}
\includegraphics[width=10cm]{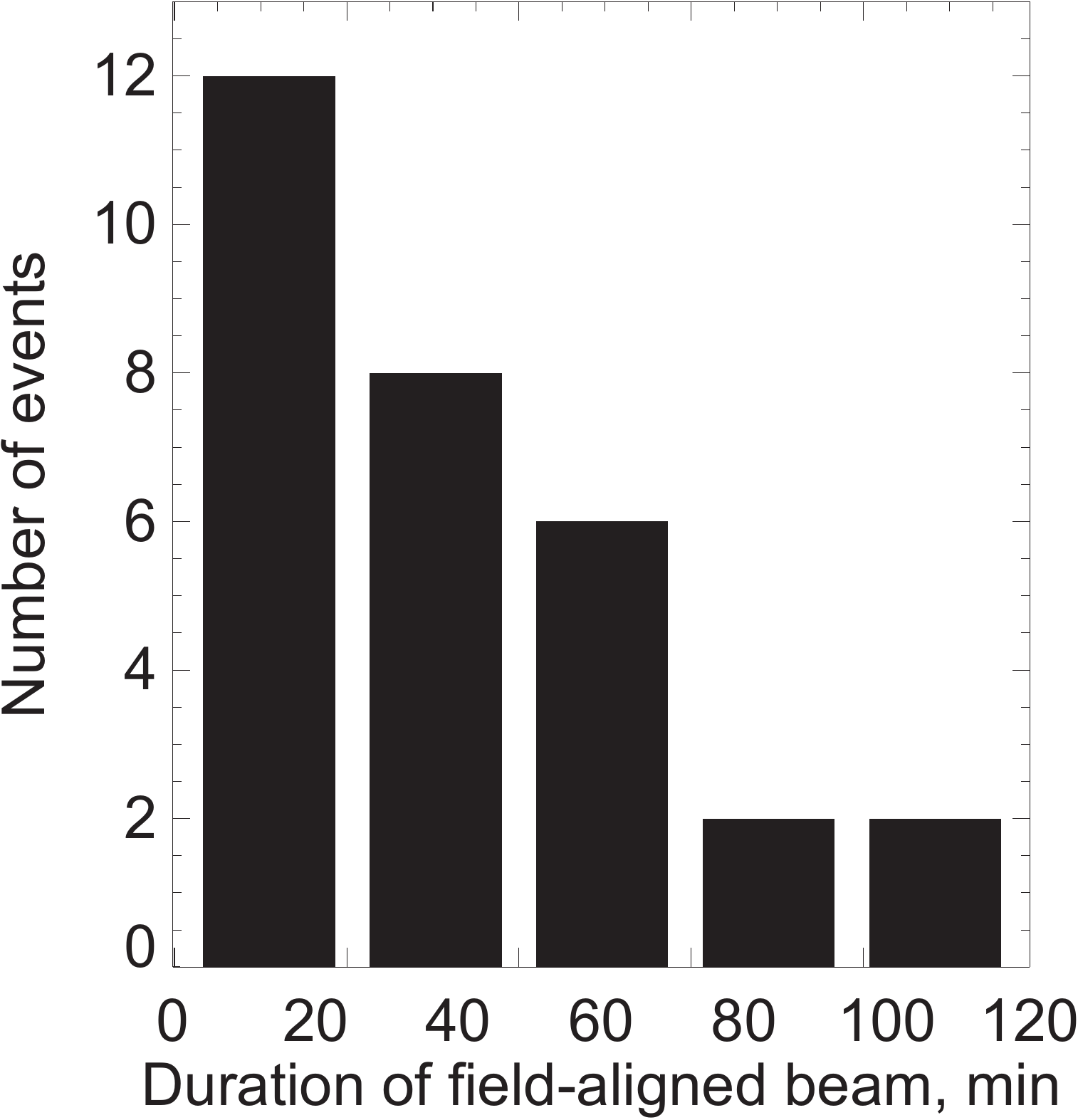}
\caption{The occurrence rate of field-aligned beams versus the duration of
the events (30 events).} \label{hist}
\end{figure}

In the previous section we presented examples of the field-aligned electron
and ion beams in energetic particle data measured onboard the Galileo
spacecraft in the jovian magnetotail. We investigated the Galileo orbits G2,
E6, G8, C9, C10 and E16, searched for such field-aligned beams and found 30
events. For these events we required:\\ (1) observation during more than one
accumulation time (that excludes spikes due to instrumental effects);\\ (2)
the radial flow anisotropy dominates the corotational one and\\ (3) the ion
pitch angle distributions show that the particle flow is predominantly
parallel/anti-parallel to the magnetic field. The distribution of these
events in the jovian magnetosphere along the Galileo orbits is shown in
Figure \ref{map}.
It is possible that due to the low time resolution of the EPD instrument many
field-aligned beams were not detected. We listed the duration of the
field-aligned beams in Table \ref{events}, which is on average 55.6 minutes.
Statistically most events have a duration between 20 and 40 minutes (see the
occurrence rate of field-aligned beams versus its duration in Figure
\ref{hist}). The distribution of events versus duration time confirms that it
is likely that many field-aligned beams have shorter duration than 20
minutes.
All 30 field-aligned beams investigated in this study are associated with
those 34 jovian reconfiguration events listed in \citet{Kron05}. During 15 of
them we observe the field-aligned beams. During 8 of these 15 events several
field-aligned beams are seen. These field-aligned beams seem to be a
constituent part of the jovian reconfiguration process. In Figure \ref{map}
we show the location of the X-lines derived by \citet{woch02} and
\citet{Vogt10}. The field-aligned beams are concentrated close to these
lines. Therefore, we believe that the field-aligned beams are signatures of
reconnection, which occur during the reconfiguration in the jovian
magnetotail.
Now we would like to estimate the mass and energy input of the field-aligned
beams in the content of the jovian energy budget. The mass which carries a
field-aligned beam can be estimated as follows:
$$m_b=n\cdot m_{O_+}\cdot V,$$
where $n$ is a typical number density in the jovian magnetotail
\citep{Frank02}. We assume that oxygen is the average ion contained in the
jovian plasma sheet. The volume, $V,$ of the field-aligned beam can be
assessed as the length of the beam $v_b\cdot t,$ where $v_b$ is the
field-aligned beam speed and $t$ is the duration, multiplied with the assumed
vertical extent 1 $R_J$ and the azimuthal extent 25 $R_J$. As average speed
of the field-aligned beam we use the speed of the ions in the plasma sheet
boundary layers, 800 km s$^{-1}$ derived by \citet{Kron08b} and as duration
we take an average duration of the field-aligned beam derived above, $\sim$60
min. Therefore, the mass release during one field-aligned beam is in the
order of 100 tons. This is lower compared to the mass of plasmoids derived by
\citet{Kron08, Bagenal11}, 800 tons and 2500 tons, respectively. The kinetic
energy of the field-aligned beam can be evaluated from the formula
$$E_{kin}=\frac{m_b \cdot v_b^2}{2}.$$ This will lead to $3\cdot10^{16}$J.
Therefore, the power of the field-aligned beam will be in the order of 9 TW, for
an averaged duration of the beam of $\sim$60 min. This power will be enough to
supply for instance jovian polar auroral emissions, see
\citep[e.g.,][]{Radioti10}.
In this paper we investigated the events in the PSBL or lobe regions.
 According to the X-type reconnection scenario by \citet{Schindler07}, we
 observe the field-aligned electron and ion beams in the PSBL. In the paper
 of \citet{Grig09}, two types of electron distribution are reported: (1) when
 the electrons are isotropic and (2) when electrons flow along the magnetic
 field lines. The first case is associated with the non-adiabatic
acceleration by the quasi-steady dawn-dusk electric field and is not
necessary associated with substorm-like activity. The beams of the second
type are usually generated close to the X-line. We check the distribution of
the particle flux in the three directions relative to the magnetic field
using the DPA: anti-parallel, DPA$<$40$^{\circ},$ parallel,
DPA$>$140$^{\circ}$ and perpendicular, 70$^{\circ}<$DPA$<$110$^{\circ}$. We
would like to note that due to the plasma corotation in the jovian plasma
sheet, many of the field-aligned flows will still have quite a substantial
corotational component. This is reflected in the pitch angle distributions as
an increased perpendicular component. 23 events with field-aligned ion beams
are associated with anisotropic electrons, i.e. which flow along the magnetic
field and 13 of them are associated with reconnection events studied by
\citet{Vogt10} derived from the magnetometer data. The characteristics of the
field-aligned ion beams are seen in Table \ref{events}. The propagation of
the ions and electrons parallel or antiparallel to the magnetic field confirm
the quadrupolar X-line configuration, see Figure \ref{scetch}. Slightly more
than half of the field-aligned beams are associated with reconnection events
observed by \citet{Vogt10}. Why not more? The reasons could be (1) different
arrival times of field-aligned beams and plasmoids; (2) the field-aligned
beams are mainly located in the lobe region where reconnection signatures
might be smoothed.
In 7 cases field-aligned ion beams are associated with the isotropic electron
distribution, see Table \ref{events}. Therefore, they might be accelerated
not at the X-line. Most of these events, namely 5, has no association with the
reconnection events from \citet{Vogt10}. This is expected.
\section{Summary}\label{summary}
(1) We presented two examples (Event 1, Event 3) of the counterstreaming
electron beams in the plasma sheet boundary layers. According to
\citet{Vog06, Asnes08} the counterstreaming electron beams occur due to the
bouncing between two mirror points on closed field lines in the plasma sheet.
They could also occur due to the travelling of electrons between multiple
X-lines.
(2) Most of the field-aligned ion events are associated with the
field-aligned electron beams which according to \citet{Grig09} are related to
electron acceleration in the X-line vicinity.
(3) The timing of the arrival of the field-aligned beams for the different
species in many events (e.g. Event 1, Event 2, Event 3) is in accordance to
the time-dispersion when the source is at one point/line \citep{Keiling04}.
Also \citet{Hill09} observed the time dispersion of the energetic ions of
different species using data from New Horizons.
(4) Observations of all types of the plasma release from plasmoid or flux
rope, its imprint in the PSBL -- TCR and dipolarization are in accordance to
the reconnection theory and all are associated with the electron and ion
field-aligned beams (Event 2, Event 3, Event 4).
(5) We show evidence that the electron beam propagates in the PSBL
 while isotropic distribution is observed around the current sheet center
 (Event 4).
(6) Observations of the flow reversals in 53$\%$ of the jovian reconfiguration
events from \citet{Kron05} imply X-line crossing, see an example in Section
\ref{scene}, Figure \ref{15}.
(7) All field-aligned beams are associated with the jovian reconfiguration
events.
Based on all these arguments we believe that reconfiguration processes in the
jovian magnetotail are associated with reconnection.
\section{Acknowledgements}
The authors thank A. Lagg for the provided EPD software and useful
discussions.
%
%
\bibliographystyle{elsarticle-harv}

\begin{table}[h]
\caption{LEMMS and CMS rate channels.}
\begin{tabular}{cccc}
\hline
 Ions (LEMMS) & & Electrons (LEMMS)&  \\
\hline
Channel&Energy range (keV)&Channel&Energy range (keV)\\
 \hline
 a0& 22-42&e0&15-29\\
a1&42-65&e1&29-42\\
a2&65-120&e2&42-55\\
a3&120-280&e3&55-93\\
a4&280-515&f0&93-188\\
a5&515-825&f1&174-304\\
a6&825-1680&f2&304-527\\
a7&1680-3200&f3&527-884\\
 \hline
 Ions (CMS)\\
 \hline
 Species&Channel&Energy range (keV/nuc)\\
 \hline
Protons&tp1&0.08-0.22 &\\
& tp2&0.22-0.54 &\\
&tp3&0.54-1.25 &\\
Oxygen+Sulfur&to1&0.012-0.026 &\\
Oxygen&to2&0.026-0.051 &\\
Sulfur&ts1&0.016-0.030 \\
&ts2&0.030-0.062 \\
 \hline
 \end{tabular}
\label{t2le}
\end{table}

\begin{table}[b]
\caption{List of the ion field-aligned beams.}
{\footnotesize
\begin{tabular}{lp{1.5cm}lp{1.5cm}p{2cm}p{4.5cm}}
\hline
Time interval, Year, DOY, UT&Radial Distance, $R_J$&LT&Duration, min& Association with the reconnection events from \citet{Vogt10}&Characteristics  \\
\hline
\multicolumn{6}{c}{\rule[-3mm]{0mm}{8mm}Ion beams associated with the anisotropic electrons as derived from tp1 channel.}\\
\hline                  \\
1996 264 1915--1955&88&0158&40& y& protons $\uparrow\downarrow$$^a$, electrons assymetric\\
1996 264 2055--2143&88&0158&48& y& protons, electrons $\uparrow\downarrow$, not clear case\\
1996 270 23218--271 0020&103&0230&62& n& protons, electrons $\uparrow\uparrow$$^b$, not clear case\\
1996 271 0945--1022&0230&103&37& n&protons $\uparrow\uparrow$, electrons bi-directional\\
1996 272 2335--273 0011&106&0238&36& y&  protons, electrons $\uparrow\uparrow$\\     
1996 275 0945--1145&110&0250&120& y& protons, electrons $\uparrow\uparrow$\\     
1996 275 1931--2045&110&0250&74& y& protons, electrons $\uparrow\uparrow$\\    
1996 277 2318--2357&112&0300&39& n& protons, electrons $\uparrow\uparrow$\\
1996 278 0125--0220&112&0300&55& n& protons, electrons $\uparrow\downarrow$\\    
1996 280 1602--1730&113&0309&73& n& protons, electrons from $\uparrow\downarrow$ to $\uparrow\uparrow$, current sheet crossing\\
1996 283 1220--1315&112&0330&50& n& protons $\uparrow\downarrow$, electrons bi-directional\\     
1996 283 2018--2117&112&0330&59& y& protons, electrons $\uparrow\downarrow$\\      
1996 284 0029--0205&112&0330&91& y&  protons $\uparrow\uparrow$, electrons bi-directional\\    
1996 285 0415--0450&111&0328&35& n & asymmetric $\uparrow\uparrow$ protons and electrons\\
1996 285 0525--0725&111&0328&120& y& protons, electrons $\uparrow\uparrow$\\    
1996 285 1415--1510&111&0328&55& y& protons, electrons $\uparrow\downarrow$\\    
1997 079 0207--0229&84&0255&22& n& asymmetric $\uparrow\uparrow$ protons and electrons, not clear case\\     
\hline\\
\end{tabular}
}
\end{table}
\newpage
\begin{table}[b]
\caption{List of the ion field-aligned beams.}
{\footnotesize
\begin{tabular}{lp{1.5cm}lp{1.5cm}p{2cm}p{4.5cm}}
\hline
Time interval, Year, DOY, UT&Radial Distance, $R_J$&LT&Duration, min& Association with the reconnection events from \citet{Vogt10}&Characteristics  \\
\hline
\multicolumn{6}{c}{\rule[-3mm]{0mm}{8mm}Ion beams associated with the anisotropic electrons as derived from tp1 channel.}\\
\hline                  \\
1997 152 0703--0740&100&0120&37& n& protons $\uparrow\downarrow$, electrons bi-directional\\    
1997 153 0620--0700&100&0120&40& y&  protons, electrons $\uparrow\downarrow$\\    
1997 158 0900--0922&97.5&0143&22& y& protons $\uparrow\uparrow$, electrons assymetric\\      
1997 161 1200--1300&94&0155&60& y& protons $\uparrow\uparrow$, electrons assymetric\\      
1997 166 2025--2115&79.5&0230&50& y&protons $\uparrow\downarrow$, electrons assymetric\\      
1997 166 2155--2227&79.5&0230&32& n&asymmetric $\uparrow\downarrow$ protons and electrons \\      
\hline
\multicolumn{3}{c}{\rule[-3mm]{0mm}{8mm}Ion beams associated with the isotropic electrons }\\
\hline
1996 262 1525--1629&79&0140&64& n& protons $\uparrow\uparrow$\\
1996 271 0233--0345&103&0230&72& y& protons $\uparrow\downarrow$\\    
1996 273 0350--0430&106&0238&35& n& protons $\uparrow\uparrow$\\
1996 275 1719--1841&110&0250&82& y& protons $\uparrow\uparrow$\\    
1996 275 2225--2338&110&0250&73& n& protons $\uparrow\uparrow$, not clear case\\
1997 161 1832--1920&94&0155&48& n& protons $\uparrow\uparrow$\\
1997 168 0720--0742&75&0240&22& n& protons $\uparrow\downarrow$\\
\hline\\
\end{tabular}
}
\label{events}$^a${$\uparrow\downarrow$ denote the anti-parallel
to the magnetic field flow.}
$^b${$\uparrow\uparrow$ denote the parallel to the magnetic
field flow.}
\end{table}
\newpage

\end{document}